\documentclass[useAMS, usenatbib]{mn2e}
\usepackage{amsmath}
\usepackage{graphicx}

\def\be{\begin{equation}}
\def\ee{\end{equation}}
\def\ba{\begin{eqnarray}}
\def\ea{\end{eqnarray}}

\def\brr{{\bf r}}
\def\bxi{{\mbox{\boldmath $\xi$}}}

\def\gtrsim{\mathrel{\hbox{\rlap{\hbox{\lower3pt\hbox{$\sim$}}}\hbox{\raise2pt\hbox{$>$}}}}}
\def\go{\mathrel{\raise.3ex\hbox{$>$}\mkern-14mu
             \lower0.6ex\hbox{$\sim$}}}
\def\lo{\mathrel{\raise.3ex\hbox{$<$}\mkern-14mu
             \lower0.6ex\hbox{$\sim$}}}

\begin{document}
\title[Tidal Dissipation in Giant Planets]
{Viscoelastic Tidal Dissipation in Giant Planets and Formation of
Hot Jupiters Through High-Eccentricity Migration}
\author[Natalia I Storch and Dong Lai]
{Natalia I Storch\thanks{Email: nis22@cornell.edu, dong@astro.cornell.edu}
and Dong Lai\\
Center for Space Research, Department of Astronomy, Cornell University, Ithaca,
NY 14853, USA\\}

\pagerange{\pageref{firstpage}--\pageref{lastpage}} \pubyear{2013}

\label{firstpage}
\maketitle

\begin{abstract}
We study the possibility of tidal dissipation in the solid cores of
giant planets and its implication for the formation of hot Jupiters
through high-eccentricity migration.  We present a general framework
by which the tidal evolution of planetary systems can be computed for
any form of tidal dissipation, characterized by the imaginary part of
the complex tidal Love number, ${\rm Im}[{\tilde k}_2(\omega)]$, as a
function of the forcing frequency $\omega$.  Using the simplest
viscoelastic dissipation model (the Maxwell model) for the rocky core
and including the effect of a nondissipative fluid envelope, we show
that with reasonable (but uncertain) physical parameters for the core
(size, viscosity and shear modulus), tidal dissipation in the core can
accommodate the tidal-Q constraint of the Solar system gas giants and
at the same time allows exoplanetary hot Jupiters to form via tidal
circularization in the high-e migration scenario.  By contrast, the
often-used weak friction theory of equilibrium tide would lead to a
discrepancy between the Solar system constraint and the amount of
dissipation necessary for high-e migration.  We also show that tidal
heating in the rocky core can lead to modest radius inflation of the
planets, particularly when the planets are in the high-eccentricity
phase ($e\sim 0.6$) during their high-e migration.  Finally, as an
interesting by-product of our study, we note that for a generic tidal
response function ${\rm Im}[{\tilde k}_2(\omega)]$, it is possible
that spin equilibrium (zero torque) can be achieved for multiple spin
frequencies (at a given $e$), and the actual pseudosynchronized spin
rate depends on the evolutionary history of the system.
\end{abstract}


\section{Introduction}

In recent years, high-eccentricity migration has emerged as one of the
dominant mechanisms responsible for the formation of hot Jupiters. In
this mechanism, a gas giant which is formed beyond the snow line is
first excited into a state of very high eccentricity ($e \go 0.9$) by
few-body interactions, either via dynamical planet-planet scatterings
(Rasio \& Ford 1996; Weidenschilling \& Marzari 1996; Zhou, Lin \& Sun~2007;
Chatterjee et al.~2008; Juric \& Tremaine 2008) or/and secular interactions between
multiple planets, or the Kozai effect induced by a distant companion (Wu
\& Murray 2003; Fabrycky \& Tremaine 2007; Wu, Murray \& Ramshhai~2007;
 Nagasawa, Ida \& Bessho~2008; Katz, Dong \& Malhotra~2011; Naoz et al. 2011,2013; Wu \& Lithwick
2011; Naoz, Farr \& Rasio 2012; see also Dawson \& Murray-Clay 2013).  Due to the high
eccentricity, the planet passes quite close to its host
star at periastron, and tidal dissipation in the planet 
extracts energy from the orbit, leading to inward migration and
circularization of the planet's orbit.

Tidal effects on the orbital evolution of binaries are often discussed
using the weak friction theory of equilibrium tides (Darwin 1879;
Alexander 1973; Hut 1981; Eggleton et al.~1998), according to which
the rate of decay of the semi-major axis ($a$) for a
pseudosynchronized planet can be written as
\begin{equation}
\left|{\dot{a}\over a}\right| = 6 k_2 \tau \left({G M_\star \over a_F^3}\right)\left({R_{\mathrm{p}}\over a_F}\right)^5 {M_\star \over M_{\mathrm{p}}}\sqrt{a_F \over a}\, F(e).
\label{eq:adotweak}
\end{equation}
Here, $M_{\mathrm{p}}$ and $R_{\mathrm{p}}$ are the mass and radius
of the planet, $M_\star$ is the mass of the central star, $a_F\equiv
a(1-e^2)$ is the final circularization radius (assuming orbital
angular momentum conservation), $k_2$ the tidal Love number, 
$\tau$ is the tidal lag time (assumed constant in the weak friction theory),
and $F(e)$ is a function of eccentricity of order $(1-10)$ and given by
$F(e) = f_1(e) - f_2^2(e)/f_5(e)$,
where $f_1$, $f_2$ and $f_5$ are given by Eq.~(11) of Hut (1981). 
By requiring that the high-e migration happens
on a timescale less than $10$ Gyr we can place a constraint on
$\tau$:
\ba
&& \left({G M_\star \over a^3_F}\right)^{1/2} \tau \gtrsim 3 \times 10^{-5}\left({a\over 5 \textrm{AU}}\right)^{1/2}\left({a_F\over 0.06 \textrm{AU}}\right)^6\nonumber\\
&&\qquad\times 
\left({M_{\star}\over M_{\sun}}\right)^{-3/2} 
{M_{\mathrm{p}} \over M_J} \left({R_{\mathrm{p}}\over R_J}\right)^{-5} \left({k_2\over 0.38}\right)^{-1}.
\label{eq:tauconstr}
\ea
Note that instead of $\tau$, tidal dissipation is often parametrized
by the tidal quality factor $Q \equiv (\tau \omega)^{-1}$, where
$\omega$ is the tidal forcing frequency. Thus, the above constraint on $\tau$ 
translates to $Q\lo 3\times 10^4$ at $\omega\sim (GM_\star/a_F^3)^{1/2}\sim 
2\pi/(5~{\rm d})$ [for the canonical parameters adopted in Eq.~(\ref{eq:tauconstr})].
A similar constraint can be obtained by integration over the planets' 
orbital evolution (e.g., Fabrycky \& Tremaine 2007; Leconte et al.~2010;
Matsumura, Peale \& Rasio~2010; Hansen 2012; Naoz et al.~2012; Socrates, Katz \& Dong~2012b).

The tidal $Q$ for Solar system giant planets can
be measured or constrained by the tidal evolution of their satellites
(Goldreich \& Soter 1966). For Jupiter, Yoder \& Peale (1981) derived
a bound $2\times 10^{-7}<k_2/Q_J<6\times 10^{-6}$
based on Io's long-term orbital evolution (particularly 
the eccentricity equilibrium), with the upper limit following
from the limited expansion of satellite orbits.
Recent analysis of the astrometric data of Galilean moons 
gave $k_2/Q_J=(1.1\pm 0.2)\times 10^{-5}$ for the current Jupiter-Io
system (Lainey et al.~2009), corresponding to $Q_J\simeq 3.5\times 10^4$
for the conventional value of the Love number 
$k_2=0.38$ (Gavrilov \& Zharkov 1977). With Jupiter's spin period
9.9~hrs and Io's orbital period 42.5~hrs, the tidal forcing frequency
on Jupiter from Io is $\omega=2\pi/(6.5~{\rm hr})$, and the tidal
lag time is then $\tau_J=(Q_J\omega)^{-1}\simeq 0.1$~s.

For Saturn, theoretical considerations based on the long-term
evolution of Mimas and other main moons (with the assumption that they
formed above the synchronous orbit 4.5 Gyr ago) lead to the constraint
$3\times 10^{-6}< k_2/Q_S< 2\times 10^{-5}$ (Sinclair 1983; Peale
1999).  However, using astrometric data spanning more than a century,
Lainey et al.~(2012) found a much larger $k_2/Q_S=(2.3\pm 0.7)\times
10^{-4}$, corresponding to $Q_S=(1-2)\times 10^3$ for 
$k_2=0.34$; they also found that $Q_S$ depends weakly on 
the tidal period in the range between $2\pi/\omega=5.8$~hrs (Rhea)
and $7.8$~hrs (Enceladus).

Assuming that extra-solar giant planets are close analogs of our own
gas giants, we can ask whether the aforementioned empirical
constraints on $k_2/Q$ for Jupiter and Saturn are compatible with the
extra-solar constraint [see Eq.~(2)]. 
The difference between the two sets of constraints is the tidal forcing
frequencies: For example, the Jupiter-Io constraint involves a single
frequency ($P=6.5$~hrs), while high-e migration involves tidal
potentials of many harmonics, all of them with periods longer than a
few days. Socrates et al.~(2012b) showed that the two
sets of constraints are incompatible with the weak friction theory (see also
Naoz et al.~2012): In order for hot Jupiters to undergo high-e migration
within the age of their host stars, their required tidal lag times must be
more than an order of magnitude larger than the Jupiter-Io
constraint.

Tidal dissipation in giant planets is complex, and depends strongly
on the internal structure of the planet, such as the stratification of
the liquid envelope and the presence and properties of a solid core.
There have been some attempts to understand the physics of tidal $Q$
in giant planets (see Ogilvie \& Lin 2004 for a review).  It has long
been known (Goldreich \& Nicholson 1977) that simple turbulent
viscosity in the fluid envelopes of giant planets is many orders of
magnitude lower than required by observations.  Ioannou \& Lindzen
(1993a,b) considered a prescribed model of Jupiter where the envelope
is not fully convective (contrary to the conventional model where the
envelope is neutrally buoyant to a high degree; see Guillot 2005) and
showed that the excitation and radiative damping of gravity waves in
the envelope provide efficient tidal dissipation only at specific
``resonant'' frequencies.  Lubow et al.~(1997) examined similar
gravity wave excitations in the radiative layer above the convective
envelope of hot Jupiters. So far the most sophisticated study of
dynamical tides in giant planets is that by Ogilvie \& Lin (2004) (see
also Goodman \& Lackner 2009; Ogilvie 2009,2013), who focused on
the tidal forcing of inertial waves (short-wavelength disturbances
restored primarily by Coriolis force) in the convective envelope of a
rotating planet [see Ivanov \& Papaloizou (2007) and Papaloizou \& Ivanov
  (2010) for highly eccentric orbits, and Wu (2005) 
for a different approach]. They showed that because of
the rocky core, the excited inertial waves are concentrated on a web
of ``rays'', leading to tidal dissipation which depends on the forcing
frequency in a highly erratic way. The tidal $Q$ obtained is typically
of order $10^{6-7}$.  It remains unclear whether this mechanism can
provide sufficient tidal dissipation compared to the observational
constraints.

The possibility of core dissipation in giant planets was first
considered by Dermott (1979) but has not received much attention
since. 
Recently, Remus et al. (2012a) showed that dissipation in the solid core 
could in principle satisfy the constraints on tidal $Q$ obtained by
Lainey et al. (2009) for Jupiter and by Lainey et al. (2012) for Saturn.
 
In this paper, we continue the study of tidal dissipation in the solid core of
giant planets and examine its consequences for the high-e migration
scenario and for the thermal evolution of hot Jupiters. 
In section 2 we present the general
tidal theory which may be used with any tidal response model. In
section 3 we discuss a simple viscoelastic tidal response model and
its range of applicability. In section 4 we use the general theory of
section 2 in conjunction with the model of section 3 to compute high-e
migration timescales and compare with the weak friction theory. We
also examine the effect of tidal heating in the core for the radius
evolution of the planets. We summarize our findings and conclude in 
section 5. 

\section{Evolution of Eccentric Systems with General Tidal Responses}

Here we formulate the tidal evolution equations for eccentric binary
systems. These equations can be applied to any tidal response
model, where the complex Love number (defined below) is an arbitrary
function of the tidal forcing frequency
[see also Efroimsky \& Makarov (2013), Mathis \& Le~Poncin-Lafitte (2009)
and Remus et al. (2012a) for similar formalisms]. 
This formulation is valid as long as the responses of the body to different 
tidal components are independent of each other.

We consider a planet of mass $M_{\mathrm{p}}$, radius $R_{\mathrm{p}}$ and rotation rate 
$\Omega_s$ (assumed to be aligned with the orbital angular momentum axis),
moving around a star (mass $M_\star$) in an eccentric orbit with 
semi-major axis $a$ and mean motion frequency $\Omega$. 
The tidal potential exerted on the planet by the star is given by
\begin{equation}
U(\brr,t)=-G M_\star \sum_{m} {W_{2m} r^2\over D(t)^3}
e^{-im\Phi(t)} Y_{2m}(\theta,\phi),
\label{Uext}
\end{equation}
where $(r,\theta,\phi)$ is the position vector (in spherical coordinates)
relative to the center of mass of the planet, 
$D(t)$ and $\Phi(t)$ are the time-dependent separation and phase of the 
orbit, and $m=0,\pm 2$, with $W_{20}=-(\pi/5)^{1/2}$ 
and $W_{2\pm 2} = (3\pi/10)^{1/2}$. 
The potential $U(\brr,t)$ can be decomposed into an infinite series 
of circular harmonics:
\begin{equation}
U(\brr,t)=- \sum_{m,N} U_{mN} r^2 Y_{2m}(\theta,\phi) e^{-i N \Omega t},
\label{Uextseries}
\end{equation}
where $N \in (-\infty,\infty)$ and
\begin{equation}
U_{mN} \equiv {GM_\star\over a^3}  W_{2m}F_{mN}(e),
\label{umn}
\end{equation}
with $F_{mN}(e)$ being the Hansen coefficient (e.g., called
$X_{2m}^N$ in Murray \& Dermott 2000), given by

\begin{equation}
F_{mN}(e) = {1\over \pi} \int^{\pi}_0 {\cos\left[N\left(\Psi - e \sin\Psi\right) - m\,\Phi(t)\right]\over{\left(1-e \cos\Psi\right)^2}} d\Psi,
\end{equation}
\noindent with
\begin{equation}
\cos\Phi(t) = {\cos\Psi - e\over 1-e\cos\Psi}.
\end{equation}

Each harmonic of the tidal potential produces a perturbative response
in the planet, expressible in terms of the Lagrangian displacement
$\bxi_{mN}$ and the Eulerian density perturbation $\delta\rho_{mN}$. 
These responses are proportional to the dimensionless ratio,
$U_{mN}/\omega_0^2=(M_\star/M_{\mathrm{p}})(R_{\mathrm{p}}/a)^3W_{2m}F_{mN}$, where
$\omega_0\equiv (GM_{\mathrm{p}}/R_{\mathrm{p}}^3)^{1/2}$ is the dynamical frequency of the planet.
Without loss of generality, we can write the tidal responses as
(see Lai 2012)
\ba
\bxi_{mN}(\brr,t) &=& {U_{mN}\over\omega_0^2} \bar{\bxi}_{mN} (r,\theta) e^{im\phi - iN\Omega t}, \label{xi} \\ 
\delta\rho_{mN}(\brr,t) &=&  {U_{mN}\over\omega_0^2} \delta\bar{\rho}_{mN} (r,\theta) e^{im\phi - iN\Omega t}, \label{drho}
\ea
with
\begin{equation}
\delta\rho_{mN}=-\nabla\cdot (\rho\bxi_{mN}).
\end{equation}
Note that $\delta\bar{\rho}_{mN}$ and $\bar{\bxi}_{mN}$ are in general
complex functions (implying that the tidal response is phased-shifted
relative to the tidal potential), and they depend on the forcing frequency
$\omega_{mN}$ of each harmonic in the rotating frame of the primary,
\begin{equation}
\omega_{mN} \equiv N \Omega - m\Omega_s.
\label{omega}
\end{equation}
Given the Eulerian density perturbation, we can obtain the
perturbation to the gravitational potential of the planet,
$\delta\Phi_{mN}$, by solving the Poisson equation, 
$\nabla^2\delta\Phi_{mN}=4\pi G \delta\rho_{mN}$. 
We define the dimensionless Love number
${\tilde k}_2^{mN}$ as the ratio of $\delta\Phi_{mN}$ and the $(mN)$-component
of the tidal potential $\left[U(\brr,t)\right]_{mN}=-r^2U_{mN}Y_{2m}(\theta,
\phi)\exp(-iN\Omega t)$, evaluated at the planet's surface:
\be
{\tilde k}_2^{mN} \equiv {\delta\Phi_{mN}\over \left[U(\brr,t)\right]_{mN}}
\Biggl|_{r=R_{\mathrm{p}}}.
\ee
Note that, just as $\delta\rho_{mN}$ is complex, so
in general is ${\tilde k}_2^{mN}$. We find that
\be
{\tilde k}_2^{mN} =  
{4\pi\over 5} {1\over M_{\mathrm{p}} R_{\mathrm{p}}^2} \int \delta\bar{\rho}_{mN}(r,\theta) 
e^{im\phi} r^2 Y^*_{2m} d^3 x.
\label{k21}
\ee

We now have all the information necessary to calculate the
time-averaged torque and energy transfer rate (from the orbit to the
planet):
\ba 
{\mathbf{T}} &=& \mathrm{Re} \Biggl\langle\,
\int\!\!d^3\!x\,\delta\rho(\brr,t)\,\brr\times \left[-\nabla
U^*(\brr,t)\right]\,\Biggr\rangle,\\
\dot{E} &=& \mathrm{Re} \Biggl\langle\,\int\!\!d^3\!x\, \rho(\brr)\,
{\partial\bxi(\brr,t)\over \partial t} \cdot \left[-\nabla
U^*(\brr,t)\right]\, \Biggr\rangle,
\ea
where $\langle~\rangle$ denotes time averaging. 
After plugging in the ansatz for $\bxi$ and $\delta\rho$
[Eqs.~(\ref{xi})-(\ref{drho})] and the expression for ${\tilde k}_2^{mN}$
[Eq.~(\ref{k21})], we find
\ba
T_z &=& {5\over 4\pi} T_0 \sum_{m,N} m\, 
\left[W_{2m}F_{mN}(e)\right]^2 \mathrm{Im}({\tilde k}_2^{mN}),\label{torque} \\
\dot{E} &=& {5\over 4\pi} T_0 \Omega \sum_{m,N} N\, 
\left[W_{2m}F_{mN}(e)\right]^2 \mathrm{Im}({\tilde k}_2^{mN}),\label{edot}
\ea
where $T_0 \equiv G \left(M_\star/a^3\right)^2 R_{\mathrm{p}}^5$. The tidal evolution
equations for the planet's spin $\Omega_s$, the orbital semi-major axis
$a$ and the eccentricity $e$ are
\ba
&& \dot\Omega_s={T_z\over I},\\
&& {\dot a\over a}=-{2a\dot E\over GM_\star M_{\mathrm{p}}},\\
&&{e\dot e\over 1-e^2}=-{a\dot E\over GM_\star M_{\mathrm{p}}}+{T_z\over L},
\ea
where $I$ is the moment of inertia of the planet and
$L=M_\star M_{\mathrm{p}}\left[Ga(1-e^2)/(M_\star+M_{\mathrm{p}})\right]^{1/2}$ is the orbital
angular momentum.

As noted before, ${\tilde k}_2^{mN}$ depends on the forcing frequency
$\omega_{mN}=N\Omega-m\Omega_s$ and physical properties of the
planet. We can write ${\tilde k}_2^{mN}={\tilde k}_2(\omega_{mN})$. 
In general, given a model for ${\tilde k}_2(\omega)$, the sum over $(mN)$
must be computed numerically. Note that 
${\rm Im}({\tilde k}_2^{mN})$ is related to the often-defined tidal quality 
factor $Q$ by 
\be
{\rm Im}({\tilde k}_2^{mN})\equiv \left({k_{2}\over Q}\right)_{mN},
\ee
with $k_{2}$ the usual (real) Love number, except that in our general case
$(k_2/Q)_{mN}$ is for a specific $(mN)$-tidal component.

In the special case of the weak friction theory of equilibrium tide
\footnote{Note that for equilibrium tides in general, the tidal
  response $\mathrm{Im}[{\tilde k}_2(\omega)]$ does not have to be a
  linear function of $\omega$ (i.e., constant lag time). For example,
  Remus et al.~(2012b) showed that for convective stars/planets,
  $\mathrm{Im}[{\tilde k}_2(\omega)]$ is independent of $\omega$ (i.e.,
  constant lag angle) when $\omega$ exceeds the convective turnover rate.},
one assumes $\mathrm{Im}[{\tilde k}_2(\omega)]=k_{2}\tau\omega$, with $k_{2}$ and
the lag time $\tau$ being independent of the frequency $\omega$.
In this case, the sum over $(mN)$ can be carried out analytically,
giving the usual expressions (see Alexander 1973, Hut 1981):
\ba
&& T_z= {3\,T_0\, \Omega\, k_2\, \tau\over
  (1-e^2)^6}\left[f_2-(1-e^2)^{3/2} f_5
  {\Omega_s\over\Omega}\right],\label{eq:torque_weak}\\
&& \dot E= {3\,T_0 \,\Omega^2 \, k_2 \, \tau \over (1-e^2)^{15/2}}
\left[f_1-(1-e^2)^{3/2} f_2 {\Omega_s\over\Omega}\right], 
\ea
\noindent where $f_1$, $f_2$, and $f_5$ are functions of eccentricity given by (Hut 1981)
\ba
f_1(e)&=& 1 + \frac{31}{2}e^2+\frac{255}{8}e^4+\frac{185}{16}e^6+\frac{25}{64}e^8, \\
f_2(e)&=& 1 + \frac{15}{2}e^2 + \frac{45}{8}e^4 + \frac{5}{16}e^6, \text{~and} \\
f_5(e)&=& 1 + 3e^2 + \frac{3}{8}e^4.
\ea

\section{Viscoelastic Dissipation in Giant Planets with Rocky Cores}

We now discuss a theoretical model of $k_2(\omega)$ for giant
planets based on viscoelastic dissipation in rocky cores.
We consider first a homogeneous solid core, and subsequently
introduce a homogeneous non-dissipative liquid envelope.

\subsection{Viscoelastic Solid Core}

The rocky/icy core of a giant planet can possess the characteristics
of both elastic solid and viscous fluid, depending on the frequency of
the imposed periodic shear stress or strain. Dissipation in rocks
arises from thermally activated creep processes associated with the
diffusion of atoms or the motion of dislocations when the rocks are
subjected to stress. We use the simplest phenomenological model, the
Maxwell model, to describe such viscoelastic materials (Turcotte \&
Schubert 2002). The model contains two free parameters, the shear
modulus (rigidity) $\mu$ and viscosity $\eta$. Other rheologies are
possible (see Henning, O'Connell \& Sasselov~2009), but contain more free parameters
and are not warranted at present given the large uncertainties associated
with the solid cores of giant planets.

The incompressible constitutive relation of a Maxwell solid core
takes the form
\be
\dot \varepsilon_{ij}={1\over 2\mu}\dot\sigma_{ij}+{1\over 2\eta}\sigma_{ij},
\ee
where $\varepsilon_{ij}$ and $\sigma_{ij}$ are strain and stress tensors,
respectively, and a dot denotes time derivative. For periodic forcing
$\varepsilon_{ij},\sigma_{ij}\propto e^{-i\omega t}$, the complex
shear modulus, $\tilde\mu\equiv \sigma_{ij}/(2\varepsilon_{ij})$,
is given by 
\be
\tilde\mu={\omega \mu\eta \over \omega\eta+i\mu} = {\mu\over{1 + i(\omega_M/\omega)}},
\ee
\noindent where the Maxwell frequency is 
\be
\omega_M\equiv \mu/\eta.
\ee
Clearly, the core behaves as an elastic solid (with $\tilde\mu\simeq\mu$)
for $\omega\gg\omega_M$, and as a viscous fluid
(with $\tilde\mu\simeq -i\omega\eta$) for $\omega\ll\omega_M$.

Consider a homogeneous rocky core (mass $M_c$, radius $R_c$ and
density $\rho_c$) with a constant $\tilde\mu$.
When the tidal forcing frequency $\omega$ is
much less than the dynamical frequency of the body, i.e.,
when $\omega\ll (GM_c/R_c^3)^{1/2}$ and $\omega\ll (\mu/\rho_c R_c^2)^{1/2}$,
the tidal Love number in the purely elastic case 
(${\rm Im}[\tilde\mu]=0$) can be obtained analytically (Love
1927). Following Remus et al. (2012a) we 
invoke the correspondence
principle (Biot 1954), which allows us to simply replace the
real shear modulus in the elastic solution by the full complex shear
modulus in order to obtain the viscoelastic solution, yielding
\begin{equation}
{\tilde k}_{2c} = {3\over 2} {1 \over 1+\bar{\mu}},
\label{eq:k2ofmu}
\end{equation}
where $\bar{\mu}$ is the body's (dimensionless) effective rigidity
\begin{equation}
\bar{\mu} \equiv \bar{\mu}_1+i\bar{\mu}_2 \equiv {19{\tilde \mu}\over 2\beta},
\label{eq:mubar}
\end{equation}
with $\beta\equiv \rho_c g_c R_c$ and $g_c=GM_c/R_c^2$.
Thus we have 
\begin{equation}
\mathrm{Im}({\tilde k}_{2c}) = 
{57\omega\eta\over 4\beta}\left[1+\left({\omega\eta\over\mu}\right)^2
\left(1+{19\mu\over 2\beta}\right)^2\right]^{-1}.
\label{eq:k2maxwell}
\end{equation}
Note that ${\rm Im}({\tilde k}_{2c})$ is a non-monotonic function of
$\omega$ (see Fig.~1, top panel). For $\omega\ll\omega_M$, we have
${\rm Im}({\tilde k}_{2c})\simeq 57\omega\eta/(4\beta)$;
for $\omega\gg\omega_M$, we have 
${\rm Im}({\tilde k}_{2c})\propto \omega^{-1}$. For a given core
model, the maximum 
\be
\mathrm{Im}({\tilde k}_{2c})_{\rm max} = {3 \hat\mu\over 4(1+\hat\mu)}
\ee
is attained at $\omega=\omega_M/(1+{\hat\mu})$, where
$\hat\mu\equiv19\mu/(2\beta)$.

\subsection{Application to a giant planet with a rocky core}

In order to apply the results of section 3.1 to a gas giant, we
introduce a non-dissipative fluid envelope on top of the rocky
body. While the fluid envelope does not, itself, dissipate energy, it
is deformed by the tidal potential and interacts with the central
solid body by exerting variable pressure on its surface, thus creating
additional stress. We consider a core of radius $R_c$ and density
$\rho_c$ within a planet of radius $R_p$, with a fluid envelope of
density $\rho_F$. 
We then use the analytical expression of Remus et al. (2012a), who used
Dermott's 1979 solution for the effect of a liquid envelope on the deformation
of an elastic core, together with the correspondence principle (Biot et al. 1954),
to calculate the resulting modified Love number of the core, 
defined as the ratio of the potential generated by the deformed core 
and the tidal potential, evaluated at the core radius ($R_c$):
\ba
\label{k2plusfluid}
&& \tilde{k}_{2c}= {1\over(B+\bar{\mu}_1)^2 +
   \bar{\mu}_2^2} \Biggl\{\Biggl[(B+\bar{\mu}_1)\left(C+{3\over 2\alpha}\bar{\mu}_1\right)
\nonumber\\
&&\qquad\quad +{3\over 2\alpha}\bar{\mu}_2^2\Biggr]-i A D \bar{\mu}_2\Biggr\},
\ea 
where (Remus et al. 2012a)
\ba
\alpha &=& 1 + \frac{5}{2} \, \frac{\rho_c}{\rho_F} \, \left( \frac{R_c}{R_p} \right)^3 \, \left( 1 - \frac{\rho_F}{\rho_c} \right) \: ,\nonumber \\
A &=& \left( 1 - \frac{\rho_F}{\rho_c} \right) \left( 1 + \frac{3}{2 \alpha} \right) \:, \label{A}\nonumber \\
B &=& 1 - \frac{\rho_F}{\rho_c} + \frac{3}{2} \frac{\rho_F}{\rho_c} \left( 1 - \frac{\rho_F}{\rho_c} \right) - \frac{9}{4 \alpha} \left( \frac{R_c}{R_p} \right)^5  \left( 1 - \frac{\rho_F}{\rho_c} \right)^2  \:,\nonumber \label{B}\\
C &=& \frac{3}{2} \, \left( 1 - \frac{\rho_F}{\rho_c} \right) \, \left( 1 - \frac{\rho_F}{\rho_c} + \frac{5}{2\alpha} \right) 
	     + \frac{9}{4\alpha} \left(\frac{R_c}{R_p}\right)^5 \, \left( 1 - \frac{\rho_F}{\rho_c} \right)^2 \:, \nonumber \\
D &=& \frac{3}{2} \, \left( 1 - \frac{\rho_F}{\rho_c} \right) \, \left[ 1 + \frac{3}{2\alpha} \, \left(\frac{R_c}{R_p}\right)^5 \right] \:.\nonumber
\ea

Since in our model, all the dissipation happens in the core, 
we then have, from section 2,
\begin{equation}
\dot{E} = {5\over 4\pi} \left(GM_\star^2R_c^5\over a^6\right)\Omega
\sum_{m,N} N\, \left[W_{2m}F_{mN}(e)\right]^2
\mathrm{Im}[\tilde{k}_{2c}^{mN}],
\end{equation}
where $\tilde{k}_{2c}^{mN}={\tilde
  k}_{2c}(N\Omega-m\Omega_s)$. However, rather than keep the explicit
dependence on $R_c$, we prefer to re-cast the equation such that all
core parameters appear in $\tilde{k}_2$ only. We write,
\begin{equation}
\dot{E} = {5\over 4\pi} \left(GM_\star^2R_p^5\over a^6\right)
\Omega \sum_{m,N} N\, \left[W_{2m}F_{mN}(e)\right]^2 \, \mathrm{Im}[\tilde{k}_2^{mN}],
\end{equation}
where
\begin{equation}
\tilde{k}_2(\omega) \equiv \left({R_c\over R_p}\right)^5 \tilde{k}_{2c}(\omega).
\end{equation}
This (complex) Love number is now, effectively, the Love number for the entire planet rather 
than for the core only.

\subsection{The specific case of Jupiter}

The size of the rocky/icy core of Jupiter is uncertain, with estimates
in the range of $\sim (0-10)M_\oplus$ (Guillot 2005) and $\sim
(14-18)M_\oplus$ (Militzer et al.~2008).  The viscous and elastic
properties of materials at the high pressure ($\sim 40$~Mbar) 
found at the center of giant planets are also poorly known. We mention
here values of $\eta$ and $\mu$ for several materials to give the
reader an idea for the range of parameter space involved. The inner
core of the Earth has a measured viscosity of $\eta\sim 10^{8 \pm 3}$
$\mathrm{bar\,\cdotp s}$ (Jeanloz 1990) and a shear modulus of
$\mu\sim 1500$ kbar, while the central pressure is $\sim 3600$ kbar
(Montagner \& Kennett 1996). In contrast, the Earth's mantle has
$\eta\sim10^{15}-10^{18}$ $\mathrm{bar\,\cdotp s}$, depending on depth
(Mitrovica \& Forte 2004), and shear modulus similar to the core. Icy
materials have $\eta\sim 10^{6}-10^{9}$~$\mathrm{bar\,\cdotp s}$, and
$\mu\sim 50$ kbar (Poirier, Sotin \& Peyronneau 1981, Goldsby \& Kohlstedt
2001). Evidently, $\eta$ in particular has a very large dynamical
range, and since very little is known about the interior of Jupiter,
all of this range is hypothetically accessible. In addition to varying
$\eta$ and $\mu$, we may also vary the size of the core $R_c$ and the
core density $\rho_c$.

\begin{figure}
\scalebox{0.55}{\rotatebox{0}{\includegraphics{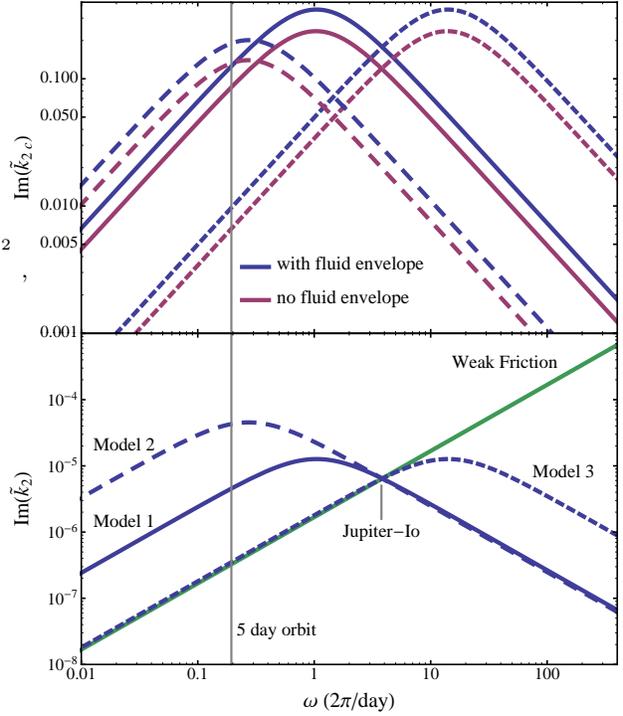}}}
\caption{Theoretical curves for the tidal Love number of a Jupiter-mass planet as a
function of the tidal forcing frequency, for several values of $R_c/R_p$ and $\eta$, each
calibrated to satisfy the Jupiter-Io constraint. 
\textit{Top:} The intrinsic Love number of the rocky core with (blue lines) and without 
(red lines) the presence of liquid envelope.
\textit{Bottom:} The effective Love number for the entire planet with fluid envelope. 
The density ratio of the core and envelope is $\rho_c/\rho_F=5$
and the core rigidity is $\mu=485\mathrm{~kbar}$ for all models. 
The other model parameters are as
  follows. \textit{Model 1 (blue solid line)}: $R_c/R_p=0.13$,
  $\eta=4.4\times10^{9}$ $\mathrm{bar\,\cdotp s}$; \textit{Model
    2 (blue long-dashed line)}: $R_c/R_p=0.19$, $\eta=2\times10^{10}$
  $\mathrm{bar\,\cdotp s}$; \textit{Model 3 (blue short-dashed line)}:
  $R_c/R_p=0.13$, $\eta=3.3\times10^{8}$ $\mathrm{bar\,\cdotp
    s}$. \textit{Green solid line}: weak friction theory with $\tau = 0.06$~s
(the lag time obtained using the value of $k_2/Q$ from Lainey et al.~(2009)
and assuming $k_2=0.38$).}
\end{figure}

Figure 1 presents three models for the tidal response
$\mathrm{Im}(\tilde{k}_{2c})$ of Jupiter's rocky core (upper panel),
and the corresponding effective tidal response of the entire planet
$\mathrm{Im}(\tilde{k}_2)$ (lower panel). For each curve, different
values of $\eta$ and $R_c$ were chosen such that the Jupiter-Io tidal
dissipation constraint is satisfied 
(see also Fig. 10 of Remus et al. 2012a). 
Also plotted is the weak friction
theory, similarly calibrated. For all the theoretical curves of Figure
1, we choose to fix $\mu$ and $\rho_c$, due to their smaller dynamical
ranges. We note that of the remaining parameters, changing $\eta$ acts
primarily to alter the transition frequency $\omega_M \sim
{\mu\eta^{-1}}$, effectively moving the curve horizontally left-right,
while changing $R_c$ effectively moves the curve up-down due to the
strong dependence of $\tilde{k}_2$ on $R_c/R_p$.

From Figure 1,  it is evident that the use of weak friction
theory, which due to having only one parameter needs only one data
point to be completely constrained, can lead to strong over- or under-
estimation of tidal dissipation at different frequencies, 
as compared to more realistic models.

\section{High-eccentricity migration of a giant planet with a rocky core}
\subsection{Orbital Evolution}

We now compute the rates of high-e migration for a giant planet with a
rocky core for different viscoelastic dissipation models depicted in
Fig.~1, and compare the results with weak friction theory. We
numerically carry out the sums in Eqs.~(\ref{torque})-(\ref{edot}) for
different values of orbital eccentricity and a fixed final semi-major
axis, i.e., the semi-major axis $a$ and eccentricity $e$ always
satisfy $a(1-e^2)\equiv a_\mathrm{F}=$constant, corresponding to a
final circular orbital period of $5$ days.

Since the timescale for changing the planet's spin is much shorter
than the orbital evolution time, we assume that the planet is in the
equilibrium spin state 
($T_z=0$) at all times. For the weak friction theory,
the result is [see Eq.~(\ref{eq:torque_weak})]
$\Omega_{ps}/\Omega_{\rm peri}=(1+e)^{-3/2}f_2/f_5$, where
$\Omega_{\rm peri}=\Omega/(1-e)^{3/2}$ is the orbital frequency at the
pericenter. For general viscoelastic models, 
we set the right-hand-side of Eq.~(\ref{torque}) to $0$ and numerically solve for
the equilibrium spin rate $\Omega_{ps}$. The results are shown in
Fig.~2.  We note that while for the model parameters considered in
Figs.~1-2, there exists a single $\Omega_{ps}$ for a given $e$ (for a
given model), as in the weak friction theory, for other model parameters 
where the torque is created by a primarily elastic rather than viscous
response ($\omega \gtrsim \omega_M$),
it is possible to find \textit{multiple} spin frequencies for which $T_z=0$, some
of which are resonant in nature, for a given $e$.  We discuss this interesting 
phenomenon in the Appendix.

\begin{figure}
\scalebox{0.55}{\rotatebox{0}{\includegraphics{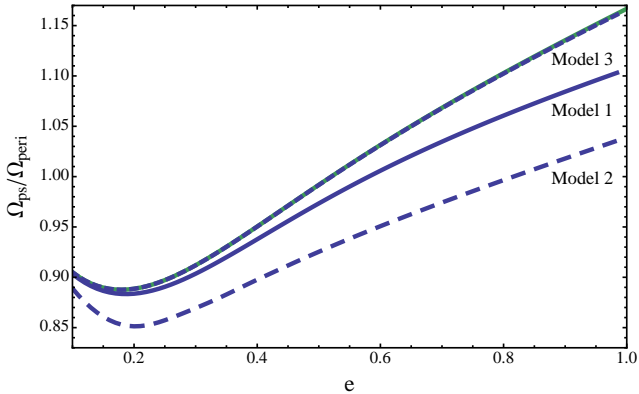}}}
\caption{Ratio of the pseudosynchronized spin frequency $\Omega_{\rm ps}$
  to the pericenter frequency $\Omega_{\rm peri}$ for each of the
  viscoelastic models of Figure 1, as well as for the (analytical)
  weak friction model. Each blue curve corresponds to one of the 
  Maxwell model curves depicted in Fig.~1. The green curve
  shows the result of the weak friction theory.}
\end{figure}

\begin{figure}
\scalebox{0.55}{\rotatebox{0}{\includegraphics{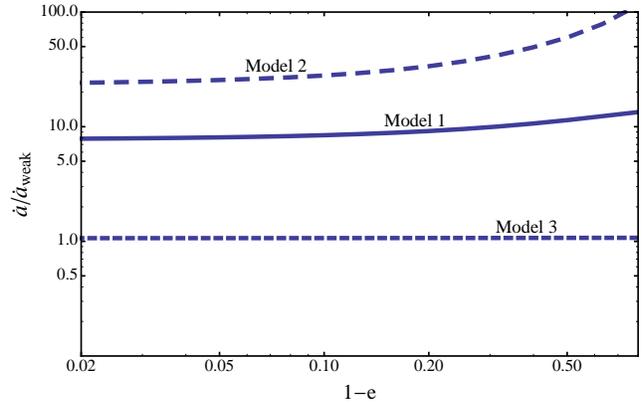}}}
\caption{Ratio of the orbital decay rate $\dot a$ for different
  viscoelastic tidal dissipation models and $\dot a_{\rm weak}$ for
  the weak friction theory, as a function of eccentricity, for a fixed
  $a_\mathrm{F}=a(1-e^2)$ corresponding to final mean motion period of $5$
  days. Each curve corresponds to one of the blue Maxwell
  model curves of Fig.~1. In all cases, the weak friction theory is
  that of the green curve in Fig.~1.}
\end{figure}

\begin{figure}
\scalebox{0.55}{\rotatebox{0}{\includegraphics{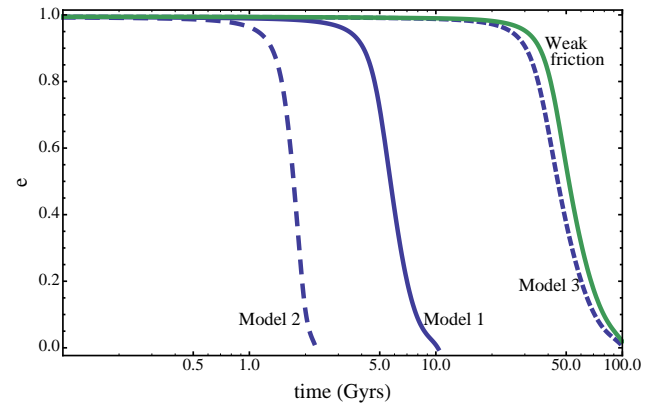}}}
\caption{Eccentricity as a function of time, for an initial eccentricity of
  $0.9945$ and a final mean motion period of $5$ days. Each blue curve
  corresponds to one of the blue theoretical Maxwell model curves of
  Figure 1. The green curve corresponds to the weak friction theory of
  Fig.~1.}
\end{figure}

Figures 3 and 4 present the results of the orbital evolution for
different viscoelastic tidal dissipation models. While all these
models satisfy the same Jupiter-Io tidal $Q$ constraint as the weak
friction theory, the predicted high-e migration rate can be
easily larger, by a factor of 10 or more, than that predicted by the
weak friction theory.  For example, while it takes $\sim 100$~Gyrs
to complete the orbital circularization in the weak friction theory, 
only $~10$~Gyrs is needed in Model 1 and only $\sim 2$~Gyrs is needed 
in Model 2. 

\subsection{Tidal heating of giant planets during migration}

Many hot Jupiters are found to have much larger radii than predictions
based on ``standard'' gas giant theory (e.g., Baraffe, Chabrier \& Barman~2010).  A
number of possible explanations for the ``radius inflation'' have been
suggested, including tidal heating (e.g., Bodenheimer, Lin \& Mardling~2001, 
Bodenheimer, Laughlin \& Lin 2003; Miller, Fortney \& Jackson~2009; Ibgui et
al.~2010; Leconte et al.~2010), the effect of thermal tides (Arras \& Socrates 2010; Socrates 2013), 
enhanced envelope opacity (Burrows et
al.~2007), double-diffusive envelope convection (Chabrier \& Baraffe 2007;
Leconte \& Chabrier 2012) and Ohmic dissipation of planetary magnetic
fields (Batygin \& Stevenson 2010; Batygin, Stevenson \& Bodenheimer~2011; but see Perna, Menou \& Rauscher~2010; Huang \& Cumming 2012; Menou 2012; Wu \& Lithwick 2013;
Rauscher \& Menou 2013).  
It is possible that more than one mechanism is needed to explain all of the
observed radius anomalies of hot Jupiters (see Fortney \& Nettelmann
2010; Spiegel \& Burrows 2013).

Several papers have already pointed out the potential importance of
tidal heating in solving the radius anomaly puzzle (see above for
references).  In particular, Leconte et al.~(2010) studied the
combined evolutions of the planet's orbit (starting from high
eccentricity) and thermal structure including tidal heating, and
showed that tidal dissipation in the planet provides a substantial
contribution to the planet's heat budget and can explain some of the
moderately bloated hot Jupiters but not the most inflated objects (see
also Miller et al.~2009; Ibgui et al.~2010). However, all these
studies were based on equilibrium tide theory with a parametrized
tidal quality factor $Q$ or lag time, and assume that the heating is
distributed uniformly across the planet.

Here we study the heating of proto-hot-Jupiters via tidal dissipation
in the core. To model this effect, we use the MESA code (Paxton et
al.~2011, 2013) to evolve the internal structure of giant planets in
conjunction with the orbital evolution starting from high
eccentricity. We create a zero-age Jupiter-mass giant planet (initially
hot and inflated) with an inert rocky core, for which we can prescribe
a time-varying luminosity. Assuming the core is in thermal equilibrium
with its surroundings, we consider the core luminosity to be equal to
$\dot{E}$ as given by Eq.~(\ref{edot}) (with $\Omega_s=\Omega_{\rm
  ps}$ such that $T_z = 0$).  We assume the planet starts at a high
eccentricity of $e=0.9945$ and circularizes to a 5-day orbit, while
conserving orbital angular momentum (so that $a_F= a(1-e^2)$ at all
times). These assumptions enable us to calculate $\dot{E}(t)$ and 
observe its effect on the radius of the planet.

Figure 5 presents the planet heating rate and radius vs age curves.
Evidently, it is possible to inflate a proto-hot-Jupiter by up to
$40\%$ via tidal heating in the core. However, this happens early in
the planet's evolution, around eccentricities of $0.6$, when the
heating rate is largest. By the time the planet's orbit has
circularized ($e\lo 0.05$), its radius is only $\sim 10\%$ larger than
the zero-temperature planet and continues to decline over time.
Therefore, regardless of the details of the tidal models, it appears
that tidal heating cannot fully explain the population of observed hot
Jupiters with significant radius inflation.  Nevertheless, tidal
effects can significantly delay the radius contraction of gas
giants. By keeping the planet somewhat inflated until (possibly)
another effect due to proximity to the host star takes over, tidal
dissipation may still play an important role in the creation of
inflated hot Jupiters.

\begin{figure}
\scalebox{0.55}{\rotatebox{0}{\includegraphics{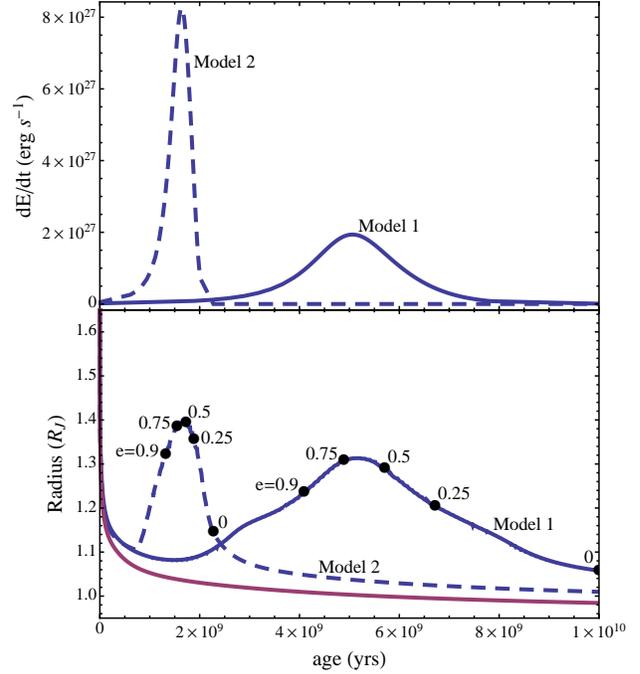}}}
\caption{\textit{Top:} Core luminosity due to viscoelastic tidal
  dissipation in a Jupiter-mass gas giant. The blue solid curve
  corresponds to Model 1 of Fig.~1, and the blue long-dashed curve
  corresponds to Model 2 of Fig.~1. \textit{Bottom:} Evolution of
  radius vs time for each of the models (top), assuming an initial
  eccentricity of $0.9945$ and a final circularized orbital period of
  5 days. The red solid curve has no tidal heating and asymptotes to
  the zero-temperature radius at later times. Black dots and labels on each curve
  denote when the particular planet model passes through that value
  of eccentricity in its orbital evolution. Note that since Model 2 
  is more dissipative than Model 1, the maximum heating rate and radius inflation 
  (around $e=0.6$) occur earlier in time than Model 1.}
\end{figure}

Interestingly, these cooling curves suggest that if tidal dissipation
in the core is indeed strong enough to play a significant role in
circularizing the planet's orbit, as we have shown to be possible in
this paper, we may expect to observe a population of gas giants (proto-hot-Jupiters) 
in wide, eccentric orbits, which are nevertheless
inflated more than expected (see Socrates et al.~2012a; Dawson \&
Murray-Clay 2013).

\section{Conclusion}

The physical mechanisms for tidal dissipations in giant planets are
uncertain. Recent works have focused on mechanisms of dissipation in the planet's
fluid envelope, but it is not clear whether they are adequate to
satisfy the constraints from the Solar system gas giants and the
formation of close-in exoplanetary systems via high-$e$ migration.  

In this paper, we have studied the possibility of tidal dissipation in
the solid cores of giant planets.  We have presented a general
framework by which the effects of tidal dissipation on the spin and
orbital evolution of planetary systems can be computed. This requires
only one input - the imaginary part of the complex tidal Love number,
${\rm Im}[\tilde{k}_2(\omega)]$, as a function of the forcing frequency
$\omega$.  We discussed the simplest model of tidal response in solids
- the Maxwell viscoelastic model, which is characterized by a
transition frequency $\omega_M$, above which the solid responds
elastically, below - viscously.  Using the Maxwell model for the
rocky/icy core, and including the effect of a non-dissipative fluid
envelope, we have demonstrated that with a modest-sized rocky core and
reasonable (but uncertain) physical core parameters, tidal dissipation
in the core can account for the Jupiter-Io tidal-$Q$ constraint (Remus et al. 2012) and at
the same time allows exoplanetary hot Jupiters to form via tidal
circularization in the high-$e$ migration scenario. By contrast, in
the often-used weak friction theory of equilibrium tide, when the
tidal lag is calibrated with the Jupiter-Io constraint, 
hot Jupiters would not be able to go through high-$e$ migration
within the lifetime of their host stars.

We have also examined the consequence of tidal heating in the rocky cores
of giant planets. Such heating can lead to modest radius inflation of the planets,
particularly when the planets are in the high-eccentricity phase ($e\sim 0.6$)
during their high-$e$ migration.

As an interesting by-product of our study, we have shown 
that when ${\rm Im}(\tilde k_2)$ exhibits nontrivial
dependence on $\omega$ (as opposed to the linear dependence in the
weak friction theory), there may exist multiple spin frequencies 
at which the torque on the planet vanishes (see Appendix A).

We emphasize that there remain large uncertainties in the physical
properties of solid cores inside giant planets, including the size,
density, composition, viscosity and elastic shear modulus. These
uncertainties make it difficult to draw any definitive conclusion
about the importance of core dissipation. Nevertheless, our study in
this paper suggests that within the range of uncertainties,
viscoelastic dissipation in the core is a possible mechanism of tidal
dissipation in giant planets and has several desirable features when
confronting the current observational constraints.  Thus, core
dissipation should be kept in mind as observations in the coming years
provide more data on tidal dissipations in giant planets.

\begin{appendix}

\section{Spin Equilibrium/Pseudosynchronization in Viscoelastic Tidal Models}

In the weak friction theory, the tidal Love number ${\rm Im}({\tilde
  k_2})$ is a linear function of the tidal frequency $\omega$, and
thus spin equilibrium
($T_z=0$) occurs at a unique value
of $\Omega_s$,
termed the pseudosynchronous frequency,
for a given orbital eccentricity $e$.  When ${\rm Im}({\tilde k_2})$ depends on $\omega$ in a more general way, as in
the case of viscoelastic tidal models of giant planets, it is possible
that multiple solutions for the equilibrium
spin frequency $\Omega_{\rm ps}$ exist at a given $e$.

\begin{figure}
\scalebox{0.55}{\rotatebox{0}{\includegraphics{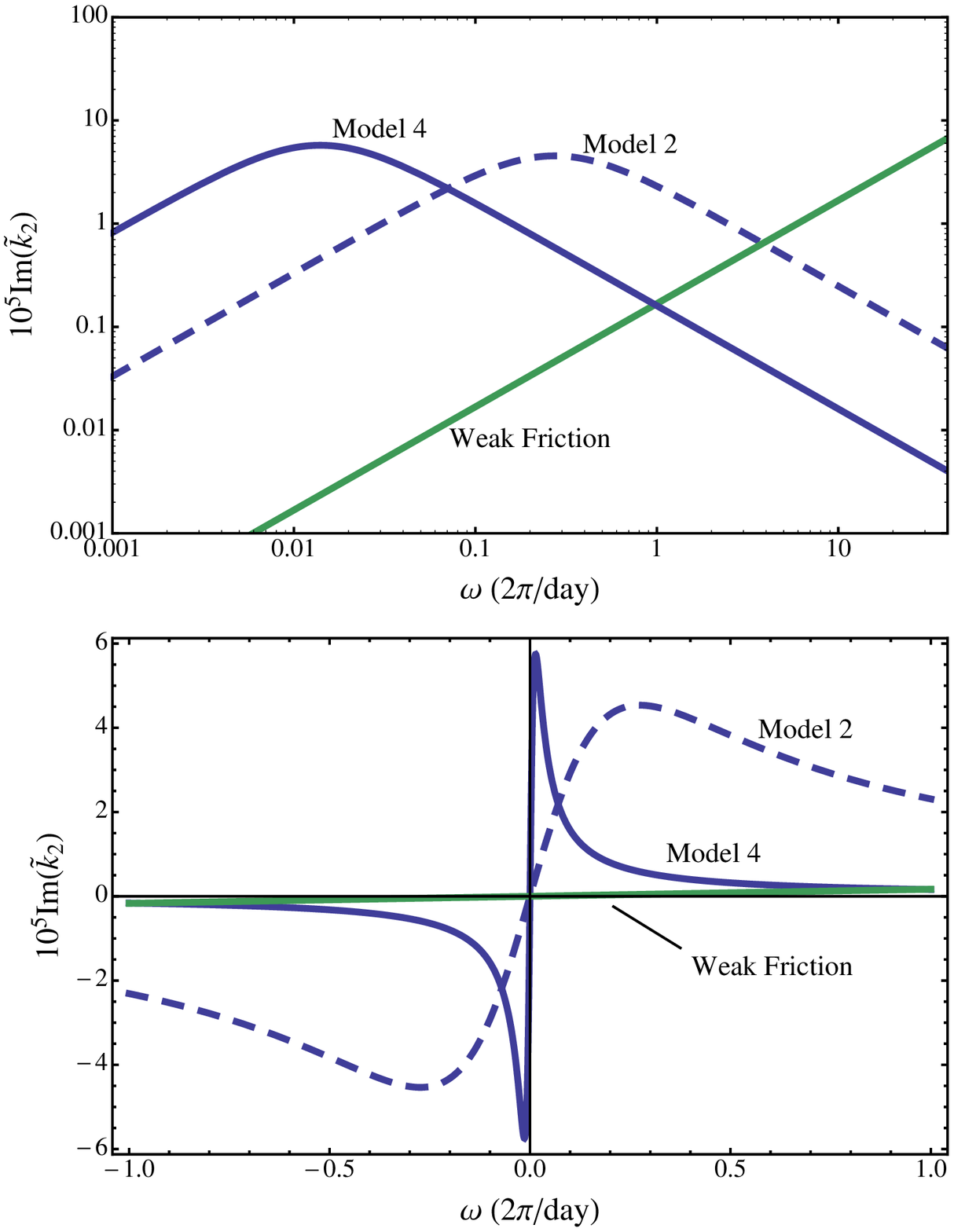}}}
\caption{Theoretical curves for the tidal Love number of a Jupiter-mass planet 
  as a function of the tidal forcing frequency. The blue dashed curve is the same as 
  Model 2 of Fig.~1, and the green solid curve corresponds to the weak friction model 
  of Fig.~1. The blue solid curve is a viscoelastic Maxwell model (model 4), 
  with $R_c/R_p=0.2$ and $\eta=4\times10^{11}$ $\mathrm{bar\,\cdotp s}$. In the top 
  panel, the models are plotted on a log-log scale, as in Fig. 1, while in the bottom panel
  we plot the models on a linear scale to clarify how the shape of the tidal response curve
  leads to resonant equilibrium spin states.}
\end{figure}

The reason for the existence of multiple pseudo-synchronized spins can
be understood in simple algebraic terms. For clarity here we demonstrate
how multiple roots arise naturally even at low eccentricities.
Consider Eq.~(\ref{torque}), which we rewrite here to make the
dependence on spin frequency explicit:
\begin{equation}
T_z = {5\over 4\pi} T_0 \sum_{m,N} m\, 
\left[W_{2m}F_{mN}(e)\right]^2 \mathrm{Im}[{\tilde k}_2(N\Omega - m\Omega_s)].
\end{equation}
For very low eccentricities $e \ll 1$, the Hansen coefficients
$F_{mN}$ are negligible for all except the following combinations of
$(m,N)$: $(0,0)$, $(0,\pm1)$,$(\pm2,\pm2)$, and $(\pm2,\pm3)$. We can
then rewrite $T_z$ as
\begin{equation}
T_z = \mathrm{Im}[A \tilde{k}_2(2\Omega-2\Omega_s) + B \tilde{k}_2(3\Omega-2\Omega_s)],
\end{equation}
with $A$ and $B$ real constants. Plugging in for $\tilde{k}_2$ using the 
Maxwell model (Eq.~\ref{eq:k2maxwell}) (neglecting fluid envelope for
simplicity), we have:
\begin{equation}
T_z = \bar{A}{(2\Omega-2\Omega_s)\over 1+ C (2\Omega-2\Omega_s)^2} +
\bar{B}{(3\Omega-2\Omega_s)\over 1+C(3\Omega-2\Omega_s)^2},
\label{toytz}
\end{equation}
where $\bar{A}$, $\bar{B}$, and $C$ are constants. Thus, when solving
for $\Omega_s$ from $T_z(\Omega_s)=0$, it is obvious that upon
finding the least common denominator, we end up solving a cubic
equation for $\Omega_s$. The above discussion can be generalized to
higher eccentricities: the pseudosynchronized spin $\Omega_{\rm ps}$
is determined by solving equations of increasingly higher (always odd)
degree in $\Omega_s$.

\begin{figure}
\scalebox{0.555}{\rotatebox{0}{\includegraphics{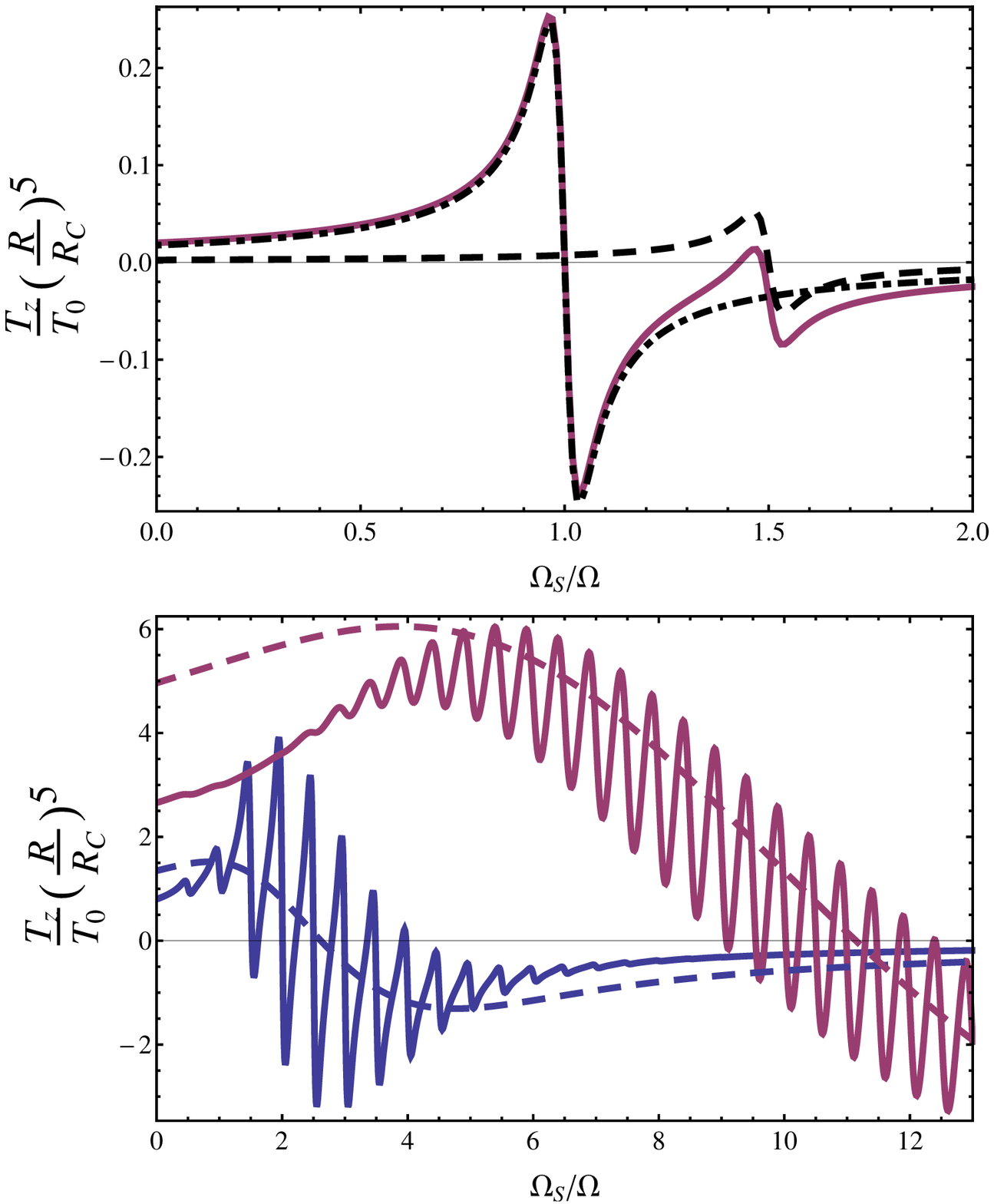}}}
\caption{Tidal torque on the planet as a function of the spin frequency for
  different values of eccentricity and different tidal dissipation
  models. \textit{Top}: Model 4 (solid blue) of Figure A1, for $e=0.13$. 
  The black dot-dashed and dashed curves show the $(m,N)=(\pm 2,\pm 2)$ and
  $(\pm 2, \pm 3)$ terms of Eq.~\ref{toytz}, respectively. The red curve
  shows their sum. The resonant features in each of the harmonics combine
  into three different zero-crossings in the sum. \textit{Bottom}: Model 4 (solid) 
  and Model 2 (dashed) of Figure A1. The red curves have $e=0.8$, 
  while the blue curves have $e=0.5$. In order to fit all the curves on the same
  plot, we show $10 T_z$ for the blue solid curve, and $0.1 T_z$ for the red
  dashed curve. Equilibrium spins are determined by $T_z=0$. Evidently, in the case
  of Model 2, the viscoelastic response is not localized enough to permit more than
  one resonant solution.}
\end{figure}

The top panel of Figure A2 shows the two terms on the RHS of Eq.~(\ref{toytz}),
as well as their sum, for the viscoelastic Model 4 of Figure A1 at an eccentricity of
$0.13$. This demonstrates
the way in which multiple solutions for $\Omega_{\mathrm{ps}}$ arise. Furthermore, we see
that two  (out of three)
of the solutions are resonant in nature: that is, they occur, roughly,
at multiples of $\Omega/2$, where $\Omega$ is the orbital frequency. This can
be understood by considering that the viscoelastic response (Figure A1) is quite
sharply peaked and localized. Each term on the RHS of Eq.~(\ref{toytz}) vanishes when
$\Omega_{\rm s}=\Omega$ and $1.5\Omega$, respectively. Due to the sharply peaked nature
of the viscoelastic response, to which $T_z$ is proportional, the sum of the two terms
then shows resonant crossings at both of these values. 

This generalizes easily to the case of arbitrary eccentricity, where 
each $(m,N)$ harmonic of the sum for $T_z$
(Eq.~A1) vanishes when $N\Omega-m\Omega_{\rm s}=0$. The number and location
of the resonant crossings then depends on the relative importance of each of 
the harmonics; the strongest crossing is expected to occur 
$\Omega_{\rm s}\sim \Omega_{\rm peri}$. This is demonstrated in Figure A2 (bottom)
 and Figure A3.

\begin{figure}
\scalebox{0.555}{\rotatebox{0}{\includegraphics{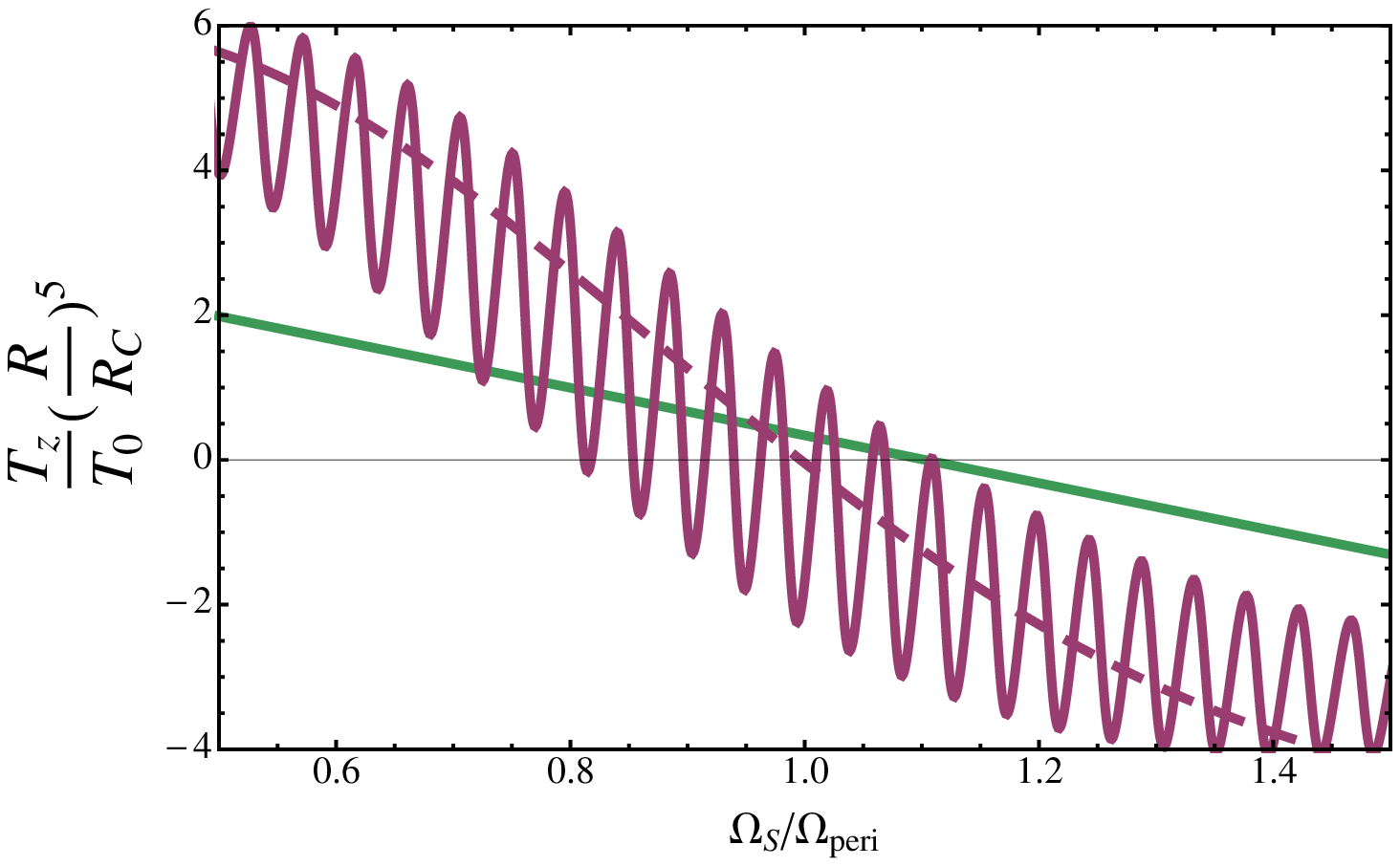}}}
\caption{Tidal torque on the planet as a function of the ratio of spin frequency to
pericenter frequency for $e=0.8$ and different tidal dissipation models.
 \textit{Red solid:} Model 4 of Figure A1; \textit{red dashed:} $0.1 T_z$ for Model 2 of Figure A1;
 \textit{green solid:} Weak friction model of Figure A1. Note that the equilibrium
 spin frequencies of the viscoelastic models can differ from the weak friction
 pseudosynchronous spin by as much as $\sim10-20\%$.}
\end{figure}

Thus, the nonlinearity of the function ${\rm Im}[\tilde k_2(\omega)]$
of the viscoelastic Maxwell model is responsible for the existence of
multiple pseudosynchronized spins.  As shown in Figure A2 (bottom
panel), there will not be multiple solutions if the viscoelastic
response is not localized enough, compared with the mean motion
frequency $\Omega$ (that is, the width of the resonant transition $\Delta
\sim \omega_M \gtrsim \Omega$), and all important harmonics of $T_z$
are solidly on the viscous (linear) side of the Maxwell curve.  On the
other hand, there may exist multiple solutions when $\Delta \sim
\omega_M \ll \Omega$ and the relevant tidal forcing frequencies lie on
the elastic side of the Maxwell curve.

Finally, we note that all the resonant zero crossings of $T_z$ are
stable (negative slope), while the non-resonant crossings are
necessarily unstable (positive slope).  The innermost and outermost
crossings are always resonant. This phenomenon is analogous to that
discussed by Makarov \& Efroimsky (2013), who used a
different model for viscoelastic dissipation in solid bodies to
analyze the pseudosynchronization of telluric planets. They
demonstrated the presence of multiple equilibrium spin solutions, and
showed that only the resonant solutions are stable equilibria, thus
concluding that the telluric planets possess no true (non-resonant)
pseudosynchronous state.

The implications of our finding may be of practical interest when it 
becomes possible to measure the spin of exoplanets on eccentric orbits. We 
may then look for evidence of the existence of multiple stable spin 
equilibria in rocky planets or gas giants with rocky cores.

\end{appendix}

\section*{Acknowledgements}

We thank 
M. Efroimsky,
M.-H. Lee, J. Lunine, P. Nicholson and Y. Wu for discussions
and information.  This work has been supported in part by NSF grants
AST-1008245, 1211061 and NASA grant NNX12AF85G.

\end{document}